\documentclass[preprint,showpacs,aps]{revtex4}
\begin{document}
\tolerance=5000
\def\be{\begin{equation}}
\def\ee{\end{equation}}
\def\bea{\begin{eqnarray}}
\def\eea{\end{eqnarray}}

\title{Towards the entropy of gravity time-dependent models via the
Cardy-Verlinde formula}

\author{Octavio Obreg\'on}
\altaffiliation[Permanent address: ]
   {Instituto de F\'\i sica de la Universidad de Guanajuato,
                    P.O. Box E-143, 37150 Le\'on Gto., M\'exico}
\email{octavio@ifug3.ugto.mx}
\affiliation{
      Department of Applied Mathematics and Theoretical Physics \\
      Wilberforce Road, Cambridge CB3 OWA, U.K.}

\author{Leonardo Pati\~no and Hernando Quevedo}
\email{jaidar, quevedo@nuclecu.unam.mx}
\affiliation{   Instituto de Ciencias Nucleares,
     Universidad Nacional Aut\'onoma de M\'exico \\
     P.O. Box 70-543, 04510 M\'exico D.F., M\'exico}

\date{\today}

\begin{abstract}
For models with several time-dependent components generalized
entropies can be defined. This is shown for the Bianchi type IX model.
We first derive the Cardy-Verlinde formula
under the assumption that the first law of thermodynamics is valid.
This leads to an explicit expression of the total entropy associated
with this type of universes. Assuming
the validity of the Cardy entropy
formula, we obtain expressions for the corresponding
Bekenstein, Bekenstein-Hawking
and Hubble entropies. We  discuss
the validity of the Cardy-Verlinde formula and possible
extensions of the outlined procedure
to other time-dependent models.
\end{abstract}

\maketitle

\section{Introduction}

The concept of entropy for cosmological models
has many subtleties. One usually starts by demanding
that the thermodynamic system described by the corresponding time-dependent
model
satisfies the first law of thermodynamics. Then, one considers a specific
equation of state and imposes compatibility between field equations and the
first law of thermodynamics. As a result one usually obtains explicit
expressions
for the thermodynamic variables (temperature and entropy)
of the model. In a further
step, the question is addressed about the physical relevance of the state
variables,
the equation of state linking them, and the thermodynamic variables. In the
case
of a simple Friedman-Robertson-Walker (FRW) universe this procedure can
successfully be performed \cite{kolb}, yielding physically reasonable
thermodynamic and state variables.

Recently, a new approach has been proposed by Verlinde \cite{ver}
who noticed a formal relationship between the field equations for a FRW
cosmology and the thermodynamic formulas of conformal field theory (CFT).
Although at first sight this seems to be only a remarkable coincidence,
a deeper analysis seems to link these results with the novel conceptual
ideas inherent in the holographic principle.

The conceptual origin of the holographic principle was settled in the 70's
by Bekenstein \cite{bek} and Hawking \cite{haw}, who formulated the second
law
of thermodynamics for black holes and the process of particle creation by
black
holes, respectively. These ideas were implemented in the context of
quantum field theories, specially in quantum gravity and cosmology by
t'Hooft \cite{thooft} and Susskind and Fischler \cite{suss,fissus}.
The holographic principle is a statement about
the counting of the quantum states of a physical system. It states
that the degrees of freedom contained in a given spatial volume can be
encoded on its boundary and that the density on the boundary does not
exceed one degree of freedom per Planck area.
This is inspired by the behavior of black hole entropy,
that the number of degrees of freedom of a gravitating system
in a region $L$ of space is the same as that of a system
on the boundary of $L$. This assumption has as a consequence also
that the number of degrees of freedom should grow only slower than
volume.
These restrictions are essentially non local and would require
ultimately profound reformulations of our description of fields
and their interactions.

The main ideas of the holographic principle have been used recently
in the context of gravity, string theory, and CFT's.
In particular, the Cardy formula \cite{Cardy}, which allows the counting
of quantum states in a two-dimensional conformal field theory, has been
generalized by Verlinde \cite{ver} to include arbitrary spacetime dimensions
and determine the density of states.
This result is now commonly referred to as the Cardy-Verlinde formula.
Verlinde also proposed that a closed universe has a Casimir contribution to
its energy and entropy and that the Casimir energy is bounded from above
by the Bekenstein-Hawking energy. Only for the case of a radiation dominated
universe this bound leads to a unification of
the Bekenstein and Hubble entropy bounds
for weakly and strongly self-gravitating universes, respectively.
Furthermore, the Cardy-Verlinde formula turns out to coincide with the
Friedman equation at the moment when the bound on the Casimir energy
becomes saturated. These results give rise to the natural question:
Is this formal merging between the Friedman formula, the Cardy entropy
and the holographic principle only a remarkable coincidence or is
it the manifestation of a profound physical property of gravity
time-dependent models,
CFT's and the holographic principle? To answer
this question it is then natural to try to generalize these fundamental
conceptions to more general physical configurations. In this context,
Wang {\it et al.} \cite{WAS}
have generalized the Cardy-Verlinde formula to the case
of universes with a cosmological constant, Nojiri {\it et al.}
\cite{quantum} have found quantum bounds for that formula, Youm \cite{youm}
has generalized the entropy formula to include perfect fluids with
a general barotropic equation of state, Brevik and Odintsov \cite{brod,brev}
considered generalizations with a constant bulk viscosity. A more extensive
list of recent works can be found in \cite{youm}.

The question we would
like to answer is if the previous results can be generalized to any
dimension and various time-dependent metric components.
In this paper, we restrict ourselves to the analysis, in Section II,
of the Bianchi type IX cosmological model
with a perfect fluid as source. As is well known, this model reduces
to the closed FRW model in the isotropic limit.
In Section III, we use the first
law of thermodynamics together with the energy conservation law
in order to obtain an expression for the entropy which turns out
to be a generalization of the Cardy-Verlinde entropy formula.
Then, in Section IV, we assume
the validity of the Cardy entropy formula
and
show that it represents the Hamiltonian constraint of the
field equations, when the Virasoro operator and the central charge
are defined in a suitable way. This allows us to propose definitions of
the
Bekenstein, Bekenstein-Hawking
and Hubble entropies, which turn out to depend explicitly on
the cosmological anisotropies.
We compare the results given in these last
two sections and  discuss
the validity of the Cardy-Verlinde formula for cosmological models
with different expansion factors. Section V is devoted to conclusions.

\section{The Bianchi type IX model}

As a particular case of a gravity time-dependent model, 
let us consider the general line element for the Bianchi type 
IX cosmological model \cite{bianchi}
\be
ds^2 = dt^2 - e^{-2\Omega} \left[ e^{2 X + 2 Y} (\omega^1)^2
                                 +e^{2 X - 2 Y} (\omega^2)^2
                                 +e^{-4 X} (\omega^3)^2 \right] \ ,
\label{bix}
\ee
where $\Omega, X,$ and $Y$ are functions of the  time $t$ only
and
\be
\omega^1 = {1\over 2}( - \sin x^3 dx^1 + \sin x^1 \cos x^3 dx^2 )\ ,
\ee
\be
\omega^2 = {1\over 2}(  \cos x^3 dx^1 + \sin x^1 \sin x^3 d x^2 )\ ,
\ee\be
\omega^3 = {1\over 2}(  \cos x^1 d x^2 + dx^3 )\ ,
\ee
where $(x^1,x^2,x^3)$ are the Euler angles of $SO(3)$.
In order to write the corresponding
field equations in a form more suitable for our
analysis,
we introduce the scale factors
\be
a_1= e^{-\Omega + X + Y} \ , \quad
a_2= e^{-\Omega + X - Y} \ , \quad
a_3= e^{-\Omega - 2 X } \ ,
\label{as}
\ee
with their corresponding Hubble parameters $H_i=\dot a_i/ a_i\ (i=1,2,3)$.
Using then
the relationships $H_1+H_2+H_3=-3\dot\Omega$, $ H_1+H_2-2H_3=6\dot X$, and
$H_1-H_2 =2\dot Y$, we can write the Hamiltonian constraint as
\be
{1\over 3}(H_1H_2 + H_1H_3 + H_2H_3)
+ {1\over a_1^2}\left(1+{1\over 3}\epsilon^2\right)  = {8\over 3}\pi G\rho \
,
\label{ham1}
\ee
with
\be
\epsilon^2 = 1 - {a_3^2\over a_2^2} - 2\left(1 -{a_1^2\over a_2^2}\right)
-{a_2^2\over a_3^2}\left(1 - {a_1^2\over a_2^2}\right)^2 \ .
\label{exe}
\ee
Furthermore, the remaining dynamical equations can be put in the form
\bea
\dot H_{1} + \dot H_{3} + H_{1}^{2}+ H_{3}^{2}+ H_{1}H_{3}
+\frac{a_{1}^{2}}{a_{2}^{2}a_{3}^{2}}-3\frac{a_{2}^{2}}{a_{1}^{2}a_{3}^{2}}
+\frac{a_{3}^{2}}{a_{1}^{2}a_{2}^{2}}+2\frac{1}{a_{1}^{2}}-2\frac{1}{a_{2}^{2}}+2\frac{1}{a_{3}^{2}}=-8\pi
G p,
\label{fe1}
\eea
\bea
\dot H_{2} + \dot H_{3} + H_{2}^{2}+ H_{3}^{2}+ H_{2}H_{3}
-3\frac{a_{1}^{2}}{a_{2}^{2}a_{3}^{2}}
+\frac{a_{2}^{2}}{a_{1}^{2}a_{3}^{2}}
+\frac{a_{3}^{2}}{a_{1}^{2}a_{2}^{2}}
-2\frac{1}{a_{1}^{2}}+2\frac{1}{a_{2}^{2}}
+2\frac{1}{a_{3}^{2}}=-8\pi G p,
\label{fe2}
\eea
\bea
\dot H_{1} + \dot H_{2} + H_{1}^{2}+ H_{2}^{2}+
H_{1}H_{2}+\frac{a_{1}^{2}}{a_{2}^{2}a_{3}^{2}}
+\frac{a_{2}^{2}}{a_{1}^{2}a_{3}^{2}}
-3\frac{a_{3}^{2}}{a_{1}^{2}a_{2}^{2}}
+2\frac{1}{a_{1}^{2}}
+2\frac{1}{a_{2}^{2}}-2\frac{1}{a_{3}^{2}}=-8\pi G p,
\label{fe3}
\eea
where $\rho$ is the matter density, $p$ its pressure, and $G$ denotes
Newton's
gravitational constant. An overdot represents differentiation with respect
to the time $t$.

From the dynamical field equations (\ref{fe1})-(\ref{fe3})
we obtain the energy-momentum
conservation law  which in this case is given by
\be
\dot\rho - 3\dot\Omega (\rho+p) = 0 \ .
\label{cons}
\ee
For a barotropic equation of state $p=\omega \rho$ the last equation
can be integrated and yields
\be
\rho =\rho_0 e^{3(1+\omega)\Omega} \ ,\
\label{rho}
\ee
where $\rho_0$ is a constant.

\section{The entropy from the first law of thermodynamics}

Let us now suppose that the Bianchi type IX model satisfies the first law
of thermodynamics, $TdS = d E + p dV$, where $T$ is the temperature of
the universe and $S$ is the total entropy contained inside the volume $V$.
For a comoving volume element of unit coordinate volume and physical volume
$v$, from the first law of thermodynamics we obtain
\be
\dot s = {v\over T}\left[\dot\rho + (p+\rho) {\dot v\over v}\right] \ ,
\label{firstlaw}
\ee
where $s$ is the entropy density per comoving volume. A reasonable way for
defining the physical volume is as $v= a_1a_2a_3=e^{-3\Omega}$,
an expression which coincides with the physical volume of the
Friedman-Robertson-Walker
universe due to the special form of the anisotropies in the Bianchi type IX
model. This assumption is supported also by the fact that the total
volume of the universe turns out to be
$V=\int \sqrt{-h} d^3x = {1\over 8} e^{-3\Omega}\int \sin x_1\, d^3x$, where
$h$ is the determinant of the spatial part of the line element (\ref{bix}).
Therefore, we can write $V=\kappa v = \kappa e^{-3\Omega}$,
where $\kappa$ is a constant resulting from the previous integral. From
Eq.(\ref{firstlaw}) and the
energy-momentum conservation law (\ref{cons}) it follows that the
entropy density remains invariant ($\dot s =0$) during the evolution, i.e.,
the expansion is adiabatic. Furthermore, the first law of thermodynamics
can be integrated for $s$ and yields \cite{kolb}
\be
s={e^{-3\Omega}\over T}(p+\rho)+s_0 \ ,
\label{entropy}
\ee
where $s_0$ is an additive constant.

Following Verlinde's approach \cite{ver} we represent the total energy
$E=\rho V$
as the sum of an extensive energy $E_E(S,V)$ and a sub-extensive energy
$E_C(S,V)$
\be
E= E_E + {1\over 2} E_C \ ,
\label{energy}
\ee
where $E_C$ is the Casimir energy which is defined as the violation
of the Euler identity:
\be
E_C = 3(E + pV - TS)\ .
\label{casimir}
\ee
Under a scale transformation $S\rightarrow \lambda S$ and
$V\rightarrow \lambda V$ the components of the energy behave as
$E_E(\lambda S, \lambda V) = \lambda E_E(S,V)$ and
$E_C(\lambda S, \lambda V) = \lambda^{1/3} E_C(S,V)$.

Introducing the expression for the entropy (\ref{entropy}) into
Eq.(\ref{casimir}) we obtain
\be
E_C = -3 {s_0\over s-s_0} (p + \rho) V \ ,
\ee
or
\be
E_C = - 3\kappa {s_0\rho_0\over s-s_0} (1+\omega) e^{3\omega \Omega} \ ,
\label{ecardy}
\ee
for a universe satisfying a barotropic equation of state. Therefore, the
behavior of the Casimir energy during the evolution in time
is given by $E_C \sim e^{3\omega\Omega}$. On
the other hand, we see from Eq.(\ref{rho}) that the total energy
of the system is given by $E=\kappa \rho_0 e^{3\omega\Omega}$
and, consequently, the extensive energy behaves similarly, $E_E \sim
e^{3\omega\Omega}$.

The total entropy $S$ of the universe should be related to the energy in
such a way that the adiabaticity condition derived above holds.
Therefore, if we associate the entropies $S_E$ and $S_C$ to the extensive
and Casimir energy, respectively, from the energy behavior given above
we conclude that only the expressions $S_E\sim E_E e^{-3\omega\Omega}$
and  $S_C\sim E_C e^{-3\omega\Omega}$ satisfy this condition.
Furthermore, taking into account the behavior of the energies
under a scale transformation one can derive their dependence
in terms of the entropy. These expressions can be written in
a convenient form as
\be
E_E = {\alpha\over 4\pi} e^{3\omega\Omega} S^{1+\omega} \ ,
\quad
E_C={\beta\over 2\pi} e^{3\omega\Omega} S^{1/3+\omega}  \ ,
\ee
where $\alpha$ and $\beta$ are arbitrary constants (In CFT's
one has that $\sqrt{\alpha\beta} = 3$ as a consequence of the
AdS/CFT correspondence). Then, the total
entropy is related to the energy as \cite{youm}
\be
S=\left[ {2\pi\over \sqrt{\alpha \beta} } e^{-3\omega\Omega}
\sqrt{2E_E E_C}\right]^{3\over 2 + 3\omega} \ .
\label{ent}
\ee
This expression for $S$ represents the generalization of the
original Cardy-Verlinde formula to the case of an anisotropic
Bianchi IX model, and reduces to it only in the case of a radiative
universe $(\omega =1/3)$ in accordance to the results in \cite{ver}.

Equation (\ref{ent}) defines the total entropy of a universe described
by  a Bianchi type IX model. It resembles the entropy for a FRW universe.
One could wonder where the time-dependent scale factors
of the Bianchi model are present
in this expression. In fact, they are not. The reason is that in the
analysis based upon the first law of thermodynamics,
the explicit form of the total volume $V$ and the physical volume $v$
is crucial and, as we have
shown, they do not depend explicitly on the functions $X$ and $Y$ which
determine the three different rates of
expansion.

Notice that for a fixed value of $E$ and positive $\omega$
the entropy (\ref{ent}) has a
maximum value so that the  bound
\be
S\leq \left[{2\pi\over\sqrt{\alpha\beta}}
  E e^{-3\omega\Omega}\right]^{3\over 2+3\omega } \ ,
\label{bekbound}
\ee
holds. This
becomes the Bekenstein bound, $ S\leq S_B$ with
$S_B = {2\pi\over 3} E e^{-\Omega}$ for a radiation dominated universe
with $\omega=1/3$ and normalization factor $\sqrt{\alpha\beta} = 3$.
It seems therefore appropriate to introduce
\be
S_B^{flt}=
\left[{2\pi\over\ 3}
  E e^{-3\omega\Omega}\right]^{3\over 2+3\omega } \ ,
\label{bekflt}
\ee
as the Bekenstein entropy which bounds the total entropy of the Bianchi type
IX
cosmological model. The superscript ``$flt$" means that this expression
for the entropy has been obtained by using the first law of thermodynamics,
and will be helpful when comparing it with a different expression
which will be derived in the next section. If we insert into
Eq.(\ref{bekflt})
the expression for the total energy, we find that $S_B^{flt}$ remains
constant during the cosmological evolution. Hence, the Bekenstein entropy
represents a constant bound for the total entropy of the system.

\section{The entropy from CFT}

The universal validity of the Cardy entropy formula
\be
S_C = 2 \pi \sqrt{ {c\over 6} \left(L_0 -{c\over 24}\right)} \ ,
\label{cardy}
\ee
for CFT's has been proposed in \cite{ver} by using a definition of
the central charge $c$ in terms of the Casimir energy.
We intentionally
use here the notation $S_C$ for the Cardy entropy in order to distinguish it
from the entropy $S$ derived in the last section by means of the first
law of thermodynamics. Let us assume
here that this entropy formula is valid for the Bianchi type IX
model. Then, independently of the equation of state
governing the evolution of
this cosmological model,
it can easily be seen that the Hamiltonian constraint (\ref{ham1}) coincides
with Eq.(\ref{cardy}) if
the eigenvalue $L_0$ of the Virasoro operator and the central charge $c$
are chosen as
\be
L_0 = {1\over 3} E {a_1\over \sqrt{1+\epsilon^2/3} }\ ,\quad
c= {3\over \pi G} {V\over a_1} \sqrt{1+\epsilon^2/3}\ ,
\label{l0c}
\ee
Furthermore, the Cardy
entropy is related to the Hubble parameters as
\be
S_C= {1\over 2\sqrt{3} G}\sqrt{ H_1H_2 + H_1H_3 + H_2H_3}\, V \ .
\label{sani}
\ee
Notice that the explicit values given in Eqs.(\ref{l0c})-(\ref{sani})
can formally be obtained by applying the transformation
\be
H \rightarrow {1\over \sqrt{3}}\sqrt{ H_1H_2 + H_1H_3 + H_2H_3} \ ,\quad
a \rightarrow {a_1\over \sqrt{1+\epsilon^2/3} } \ ,
\label{trans}
\ee
on the Hubble parameter $H$ and the scale factor $a$ of the isotropic
FRW spacetime. Written in this form, this transformation seems to
indicate the presence of a preferred direction determined by an
inhomogeneity in $a_1$. However, similar expressions can be written
for $a_2$ and $a_3$ (with the corresponding changes in the definition
of $\epsilon^2$) and, therefore any of the directions could have been
chosen.
The specific direction chosen in order to perform the transformation
(\ref{trans})
is only a matter of conventions.
Also, it should be
mentioned that once we assume that the Cardy formula (\ref{cardy}) is
valid for the Bianchi type IX model, and we demand that it represents
the field equation (\ref{ham1}), then the identification of $L_0$ and $c$
given in (\ref{l0c}) is unique (modulo the convention mentioned above).

In a similar manner, the Hamiltonian constraint (\ref{ham1})
can be rewritten as the Verlinde formula
\be
S_H^2 - 2 S_{BH} S_B + S_{BH}^2 = 0 \ ,
\label{verlinde}
\ee
where the Bekenstein, $S_B$, the Hubble, $S_H$,
and the Bekenstein-Hawking entropy, $S_{BH}$, are given by
\be
S_B = {2\pi \over 3} {E a_1\over \sqrt{1+\epsilon^2/3} } \ ,
\label{sb}
\ee
\be
S_{BH} = {1\over 2 G} {V\over a_1} \sqrt{1+\epsilon^2/3} \ ,
\label{sbh}
\ee
\be
S_H = {1\over 2\sqrt{3} G} \sqrt{ H_1H_2 + H_1H_3 + H_2H_3}\, V \ .
\label{sh}
\ee
Then, by analogy with the cases studied in other works (see \cite{youm}
for a list of references)
in which the relationship
between the Cardy formula and the field equations has been investigated,
we propose expressions (\ref{sb})-(\ref{sh}) as the definitions
of the entropies for the Bianchi type IX model. In the isotropic
limit ($\epsilon^2 =0)$ we recover the corresponding entropies
for the FRW model.

Eq.(\ref{verlinde}) has exactly the same Pythagorean form as that
found for the FRW model \cite{ver}.
This allows us to represent the dynamical
evolution of the entropies by a circle of radius $S_B$ which, in contrast
to the FRW case, is not constant but changes as time evolves.
As a consequence, the dynamical evolution of any of the entropies defined
in Eqs.(\ref{sb})-(\ref{sh}), which is
governed by the field equations (\ref{fe1})-(\ref{fe3}), is no longer
determined by the remaining entropies but depends on the specific
characteristics of the fluid. For instance, from Eqs.(\ref{sb}) and
(\ref{sbh})
and using the field equations (\ref{fe1})-(\ref{fe3}) in the form
(\ref{cons})
one can show that
\be
{d\over dt}(S_B S_{BH})  = 3(\omega -1) \dot\Omega S_B S_{BH} \ .
\ee
On the other hand, from Eq.(\ref{verlinde}) we obtain
\be
\dot S_H = {S_{BH}\over S_H} [ 3(\omega -1) \dot\Omega S_B - \dot S_{BH} ] \
.
\ee
The last two equations determine the dynamical behavior of the entropies
which will then depend on the specific value of $\omega$. The
case $\omega =1$ can easily be solved, however it is not of particular
interest in this work.

To get some insight
into the structure of these entropies, one can calculate the
approximate contribution for the special
case in which one has
only two different time-dependent metric components. This can be reached
by considering $X$ as an infinitesimal
quantity ($X<<1$) and assuming $Y=0$. Then we have
\be
S_B^a = S_B^i \left(1+{5\over 2} X^2\right) \ ,
\label{sb1}
\ee
\be
S_{BH}^a = S_{BH}^i \left(1-{5\over 2} X^2\right) \ ,
\label{sbh1}
\ee
\be
S_H^a = S_H^i \left(1 - {\dot X ^2\over 2 H^2} \right) \ ,
\label{sh1}
\ee
where the superscripts $a$ and $i$ denote the corresponding `anisotropic''
and `isotropic'' quantities, respectively.

The first thing one can notice is that, due to the anisotropies,
the Bekenstein entropy given in Eq.(\ref{sb})  does not remain
constant during the evolution in time [cf. also Eq.(\ref{sb1})].
This differs from the result obtained in the last section for the
Bekenstein entropy $S_B^{flt}$ which is a constant that only depends
on the type of fluid (i.e., it depends on $\omega$). Nevertheless,
both entropies (\ref{bekflt}) and (\ref{sb}) reduce in the limiting
case $X=0,\ Y=0,\ \omega=1/3$ to the Bekenstein entropy of a radiation
dominated FRW universe.

In terms of the Hubble entropy, $S_H$, and the Bekenstein-Hawking energy,
$E_{BH}$, which are defined as
\be
S_H = S_C,
\qquad E_{BH} = {3\over 4\pi G} {V\over a_1^2}(1+\epsilon^2/3)
\ ,
\label{ebh}
\ee
Eq.(\ref{cardy}) for the Bianchi type IX model can be identified
with the cosmological Cardy formula
\be
S_H = {2\pi\over 3} {a_1\over \sqrt{1+\epsilon^2/3}}
\sqrt{E_{BH}(2E - E_{BH})} \ .
\label{coscardy}
\ee
In a radiation dominated FRW universe the Cardy-Verlinde formula
(\ref{ent}) coincides with the cosmological Cardy formula
(\ref{coscardy}), with $E_{BH}$ playing the role of the Casimir
energy $E_C$. This is not true in our general case. Whereas
the functional dependence in the cosmological Cardy formula
(\ref{coscardy}) is dictated by the square root of the energies,
independently of the equation of state, this functional dependence
in the Cardy-Verlinde formula (\ref{ent}) is different for each
equation of state and becomes a square root only for $\omega=1/3$.
Let us therefore consider only the case of a radiation dominated
Bianchi IX universe, for which the Cardy-Verlinde formula (\ref{ent})
reads (we take $\sqrt{\alpha\beta} = 3$
and use the relationship $E_E=(2E - E_C)/2$)
\be
S={2\pi\over 3  } e^{-\Omega}
\sqrt{E_C (2E-E_C)} \ ,
\label{carver}
\ee
and for this case let us investigate
the cosmological bound on the Casimir energy
\be
E_C \leq E_{BH} \
\label{ebound}
\ee
postulated by Verlinde \cite{ver}. From Eq.(\ref{ecardy})
we see that $E_C \sim e^\Omega$
while the Bekenstein-Hawking entropy behaves like
$E_{BH} \sim e^{-\Omega-2X-2Y}(1+\epsilon^2/3)$, according to
Eq.(\ref{ebh}).
Thus, the Verlinde energy bound  (\ref{ebound}) can be satisfied,
but a difference appears in the saturation. Consider,
for instance, the approximate case ($Y=0, \ X<<1$). Using the
approximate expression (\ref{sbh1}) we get
$E_{BH}^a= (3/4\pi G) e^{-\Omega}(1-8X^2) = E_{BH}^i (1-8X^2)< E_{BH}^i$. On
the
other hand, according to Eq.(\ref{ecardy}) $E_C^a = E_C^i$. Since
the bound $E_C^i\leq E_{BH}^i$ holds and becomes saturated for a specific
size of an expanding universe, then the bound
$E_C^a\leq E_{BH}^a$ will be the generalization of the isotropic case.
However, because $E_{BH}^a =E_{BH}^i (1-8X^2)$, the previous relation,
for the anisotropic case means that the bound would become
saturated for a larger size, compared with the isotropic case,
of the expanding universe.
At the saturation point $E_C^a=E_{BH}^a$, however,
the Cardy-Verlinde
formula (\ref{carver}) does not coincide with the cosmological Cardy formula
(\ref{coscardy}), because of the presence of the various scale
factors entering the last formula. In
fact, for the approximate case we obtain that $S_H = S (1+5/2 X^2)$ and
the coincidence breaks down.

\section{Conclusions}

In this work we have pursued the
definition of the entropies associated with gravity time-dependent models.
These entropies have been found for the Bianchi type IX model.
Since this model is considered the
anisotropic generalization of the closed FRW model, it should give us the
modified entropies due to the presence of several
time-dependent metric components. First, we assumed that
the universes described
by this model satisfy the
first law of thermodynamics. The resulting entropy is given in
Eq.(\ref{ent}) and represents a generalization of the Cardy-Verlinde
formula for these models. Remarkably,
this entropy remains constant during the evolution of time as a
consequence of the energy conservation law and the assumed first
law of thermodynamics.

Secondly, we assume the universal
validity of the Cardy entropy formula and show that it coincides
with the Hamiltonian constraint of the field equations. This allows
us to consider the cosmological Cardy formula (\ref{coscardy})
and define the Bekenstein, Bekenstein-Hawking and Hubble entropies.
The first obvious consequence is that these entropies
do depend on the different metric components.

The straightforward generalization of the Verlinde energy bound
for the anisotropic radiative case ($E_C^a \leq E_{BH}^a$) indicates that
the saturation takes place for a universe larger than the one
in the corresponding isotropic case. We also have shown that
the Cardy-Verlinde formula (\ref{carver})
and the cosmological Cardy formula (\ref{coscardy}) do not coincide
when the bound is saturated $(E_C^a = E_{BH}^a)$.

An elegant AdS/CFT prescription has been proposed \cite{sav} (see also
\cite{fquev} for a review) in which starting
with a Schwarzschild AdS$_5$ metric and specifying the location of a brane
in
parametric form one can induce a 4-dimensional  FRW radiation universe
metric on the brane.
In \cite{youm1} and \cite{youm2}
this approach has been generalized to include the cases of asymptotically
flat and asymptotically AdS charged black holes. This has shed new light on
Verlinde's proposal and on the coincidence between the CFT entropy
formulae and the FRW equations at the horizon.
The case of models with various scale factors
has been less explored.
Nevertheless, a geometric braneworld description of the Bianchi VI$_{-1}$
model was obtained in \cite{wool} from other solutions found to the
vacuum Einstein equations with negative cosmological constant
in 5-dimensions. However, generalized
entropies have not been investigated and it is not clear if
they could be defined for this case. On the other hand, as mentioned
above, the Bianchi IX model contains the closed FRW model in the
isotropic limit and one can
study the influence of several metric components on the entropies.
Nonetheless,
the search for a higher dimensional static
metric by means of which one could get
the Bianchi IX model on the brane is a matter of future work.

On the other hand, in this work we have analyzed
only the Bianchi type IX model.
It seems reasonable to expect that the results
obtained here can be generalized to include other Bianchi type A models
\cite{mike}.
Moreover, the results presented here can, in
principle, be generalized to any number of dimensions with several
time-dependent metric components. However, the entropy expresions 
(\ref{ent})
and (\ref{bekflt})
reduce to the original Cardy-Verlinde formula only for a radiative
universe. A next step would be to try to extend the kind of results
obtained here for various gravity time-dependent components to their 
study in relation with more general
matter fields
\cite{quantum,youm,brod,brev}.
This would allow the study of other physical interesting systems
like SD-branes \cite{gutstro}, the kind of objects that are constructed by
means of
time-dependent metric components and that arise naturally in string
theory.

\begin{acknowledgments}
We would like to thank M. P. Ryan for helpful discussions.
  This work was supported  by DGAPA-UNAM grant IN112401,  CONACyT-Mexico
  grants 36581-E,  37851-E, and sabbatical grant 020331 (O. O.).
L. P. was supported by a UNAM-DGEP Graduate Fellowship.

\end{acknowledgments}


\begin{thebibliography}{999}

\bibitem{kolb} E. Kolb and M. Turner, {\it The Early Universe}
(Addison-Wesley, 1990).

\bibitem{ver} E. Verlinde, {\em On the holographic principle in a radiation
dominated universe}, hep-th/0008140.

\bibitem{bek} J. D. Bekenstein, Lett. Nuovo Cim. {\bf 4} 737 (1972).

\bibitem{haw} S. Hawking, Commun. Math. Phys. {\bf 43} 199 (1975).
\bibitem{thooft} G. t'Hooft,
in: {\em Salamfestschrift: A collection of talks}, edited by
A. Ali, J. Ellis, and S. Randjbar-Daemi (World Scientific, Singapore, 1993),
gr-qc/9310026.
\bibitem{suss} L. Susskind, J. Math. Phys. {\bf 36} 6377 (1995).
\bibitem{fissus} W. Fischler and L. Susskind, {\em Holography and
cosmology},
hep-th/9806039.
\bibitem{Cardy} J. L. Cardy, Nucl. Phys. B {\bf 270} 967
(1986).
\bibitem{WAS} E. Wang, E. Abdalla and R. Su, Phys. Lett. B {\bf 503}
394 (2001). 
\bibitem{quantum} O. Nojiri, S. Odintsov, O. Obregon,
H. Quevedo and M.  Ryan,
Mod. Phys. Lett. A {\bf 16} 1181 (2001).
\bibitem{youm} D. Youm, Phys. Lett. B {\bf 531}, 276 (2002).
\bibitem{brod} I. Brevik and S. D. Odintsov, Phys. Rev. D {\bf 65} 067302
(2002).
\bibitem{brev} I. Brevik, Phys. Rev. D {\bf 65} 127302 (2002).
\bibitem{bianchi} C. W. Misner, K. S. Thorne and J. A. Wheeler, {\it
Gravitation}
(W. H. Freeman and Company, San Francisco, 1973).
\bibitem{sav} I. Savonije and E. Verlinde , Phys. Lett. B {\bf 507} 305
(2001).
\bibitem{fquev} F. Quevedo, Class. Quant. Grav. {\bf 19} 5721 (2002).
\bibitem{youm1} D. Youm, Mod. Phys. Lett. A {\bf 16} 1263 (2001).
\bibitem{youm2} D. Youm, Mod. Phys. Lett. A {\bf 16} 1327 (2001).
\bibitem{wool} C. Cadeau and E. Woolgar. Class. Quant. Grav. {\bf 18} 527
(2001).
\bibitem{mike} M. P. Ryan and  L. C. Shepley, {\it Homogeneous relativistic
cosmologies} (Princeton University Press, Princeton, 1975).
\bibitem{gutstro} M. Gutperle and A. Strominger,
J. High Energy Phys. JHEP 0204018 (2002); see also
A. Sen, J. High Energy Phys. JHEP 0204048 (2002),
F. Leblond and A. W. Peet, {\em SD-brane gravity fields and rolling
tachyons},
hep-th/0303035, and references therein.


\end{thebibliography}

\end{document}